\documentstyle[aps,prl,twocolumn,floats,epsf]{revtex}
\begin{document}

\def\e{\epsilon}
\def\iwn{i\omega_n}
\def\iwm{i\omega_m}
\def\ivn{i\nu_n}
\def\ul#1{\underline{#1}}
\def\mat#1{\underline{\underline{#1}}}
\def\k{\ul{k}}
\def\q{\ul{q}}
\def\dps{\displaystyle}

%%\baselineskip 19pt
%%2col
\twocolumn[
\hsize\textwidth\columnwidth\hsize\csname @twocolumnfalse\endcsname
%%%%%%%%%%%%%%%%%%%%%%%%%%
%% start of wide text
%%2col

\title{From ferromagnetism to spin-density wave: Magnetism in
       the two channel periodic Anderson model}
\bigskip
\author{Frithjof B. Anders}
\address{Institut f\"ur Festk\"orperphysik, Technical
             University Darmstadt, 64289 Darmstadt, Germany}
%%\date{\today}
\date{May 18, 1999}
\maketitle

\begin{abstract}
The magnetic properties of the two-channel periodic Anderson model for
uranium ions, comprised of a quadrupolar and a magnetic doublet are
investigated through the  crossover from the mixed-valent   to the
stable moment regime using dynamical mean field theory.  
In the mixed-valent regime ferromagnetism is found for low carrier
concentration on a  hyper-cubic lattice. The Kondo regime is governed by
band magnetism with small effective moments and an ordering vector
$\q$ close to the perfect nesting vector. In the stable moment regime nearest neighbour
anti-ferromagnetism  dominates for less than half  band filling and a  spin density wave
transition for larger than half filling.
$T_m$ is governed by the renormalized RKKY energy scale
$\mu_{eff}^2<\!\!X_\sigma\!\!>^2J^2\rho_0(\mu)$.

\end{abstract}

\bigskip
%% 71.27.+a -- Strongly correlated electron systems; heavy fermions.
%% 75.30.Mb -- Valence fluctuation, Kondo lattice, and heavy-fermion phenomena.
%% 75.20.Hr -- Local moment in compounds and alloys; Kondo effect, valence
%%             fluctuations, heavy fermions.

%%2col
%% end of wide text
]

\narrowtext
%%2col

\draft

%%%%%%%%%%%%%%%%%%%%%%%%%%%
%% Introduction
%%%%%%%%%%%%%%%%%%%%%%%%%%%
\paragraph*{Introduction.}
Heavy Fermion (HF) materials \cite{GreweSte91} have drawn 
much attention since the discovery of
superconductivity in CeCu$_2$Si$_2$ \cite{Steglich79} which is
characterized by an anisotropic order parameter with  symmetry yet  to be
determined. Over the years, it became clear that the heavy
Fermi-liquid is not the ground state: almost all  HF materials show
magnetic or superconducting phase transitions, which either compete
with each other in Ce based compounds or coexist in Uranium based
materials. Additionally,  doubts were casted on whether the
Fermi-liquid picture is even applicable for the normal state
properties in some U-based \cite{Cox87} or doped compounds
\cite{Maple95}.

In this paper, new results on magnetism in the two-channel periodic Anderson
 model (TCPAM) \cite{AndersJarCox97} are presented. In the TCPAM, the
restricted local Hilbert space, consisting of a
quadrupolar and a magnetic  ground state doublet for two different
valence configurations, models an Uranium ion in a cubic environment,
which can either fluctuate between $5f^2$/$5f^3$ or $5f^1$/$5f^2$
ground state configurations. 
On symmetry grounds, the coupling of the ion to the
itinerant electrons via hybridization can be characterized by an
orbital and a spin label  generating an effective two band 
model\cite{Cox87}.

It has been shown that the paramagnetic phase exhibits
non-Fermi-liquid properties, as manifested  in a
large residual resistivity \cite{AndersJarCox97}. 
They are caused by the over-compensated effective impurity on each
Uranium site acting as incoherent scatterer.
The predicted absence of a Drude peak in the
optical conductivity\cite{AndersJarCox97} has been found
experimentally in UBe$_{13}$ \cite{Bommeli97}. 
On a lattice, however, the residual entropy\cite{CoxZawa98} is
expected to be quenched by a phase transition. Candidates are 
magnetic, channel ordering or superconducting transitions, which
all have to bind the residual magnetic or quadrupolar spin.
In the periodic two-channel Kondo model, anti-ferromagnetic and charge 
ordering phase transitions have been reported 
 in the strong coupling limit of the model
$|J|\rho_0(0) > 0.5$\cite{JarrellPangCox97}, $\rho_0(\e)$ being
the non-interacting density of 
states. Additionally, the authors interpret a
sign change in the pair susceptibility for a spin/channel singlet
sector as a superconducting transition. The absence of a divergence 
is viewed as indicator of a first order transition.

In this paper,  I focus on  possible spin  
transitions. 
I have calculated the spin phase transition
temperature $T_m$ as a function of band filling and energy difference
$\e_f = E_\sigma -E_\alpha$ between the quadrupolar and the magnetic
doublet using the dynamical mean field theory\cite{Georges96}. 
Varying $\e_f$ continuously allows me to investigate three different
regimes:  $|\e_f|$ smaller than the charge fluctuation energy $\Delta$ 
defines the intermediate valence regime, which has recently been
discussed  experimentally \cite{Aliev95,Aliev96} and theoretically
\cite{SchillerAndCox98,KogaCox99} as a possible scenario  for
U$_{1-x}$Th$_x$Be$_{13}$. The local moments are not
well defined, and the carriers rather mobile. Magnetic ordering would
have to involve a gain in  kinetic energy since the effective exchange
couplings are small\cite{GreweKei81}. Ferromagnetism is  found
at low carrier concentration.
For $|\e_f/\Delta|\approx 2$, the Kondo 
effect dominates. Band magnetism with very small effective magnetic 
moments is expected since all available electrons are involved in the
screening process. 
For large $|\e_f/\Delta|$, magnetism is governed by
stable local moments whose interactions are mediated by the conduction 
electrons (RKKY interaction).

%%%%%%%%%%%%%%%%%%%%%%%%%%%
%% Theory
%%%%%%%%%%%%%%%%%%%%%%%%%%%

\paragraph*{Theory.}
The two channel Anderson lattice Hamiltonian under investigation reads
\begin{eqnarray}
\label{equ-hamil}
  \hat H  &= &\dps
\sum_{\alpha <i,j>} \frac{t^*}{\sqrt{d}}
c^\dagger_{i\alpha\sigma}c_{j\alpha\sigma}
+\sum_{i\sigma} E_\sigma X_{\sigma,\sigma}^{(i)}
+\sum_{i\alpha} E_\alpha X_{\alpha,\alpha}^{(i)}
\nonumber\\  &   &\dps
 +
\sum_{i\sigma\alpha} V\left\{
c^\dagger_{i\alpha\sigma}  X_{-\alpha,\sigma}^{(i)}
+ h.c
 \right\} \;\; .
\end{eqnarray}
$X$ are the  Hubbard operators,
$d$ being the spatial dimension, and $t^*$ the
reduced hopping matrix
element of the conduction electrons between nearest neighbours. The
$c$-electrons are created
by $c^\dagger_{i\alpha\sigma}$ at the lattice site  $i$, labeled by
a spin $\sigma$ and a channel index $\alpha=\pm$ and couple via the
hybridization matrix element $V$ to the ionic 
many-body states on each lattice site. We will assume that the energies of the
magnetic  doublets split linearly under an applied small external
field, and van Vleck contributions are neglected. 

In the limit of infinite spatial dimensions the two particle-hole
irreducible vertex $\Gamma_{ff}$ becomes local  \cite{BrandMilsch89}. Using
the dynamical mean field theory  (DMFT) \cite{Georges96} in finite dimensions,
the Bethe-Salpeter equation for the full two
particle-hole  propagator in 
f-electrons reads
\begin{equation}
\label{equ-bethe-salp}
\mat{\Pi}_{ff}(\ivn) = \beta \mat{\chi}^{0}_{ff} + 
\mat{\chi}^{0}_{ff}\frac{1}{\beta}\mat{\Gamma}_{ff}\mat{\Pi}_{ff}(\ivn)
\;\; ,
\end{equation}
where $\chi^{0}_{ff}$  is the free particle-hole $ff$-propagator
\begin{equation}
\label{equ-3}
\begin{array}[b]{l}
\chi^{0}_{ff}(\iwn,\iwm;\ivn,\q) = -\frac{\delta_{n,m}}{N}\sum_{\k}
F_{\k}(\iwn+\ivn) F_{\k+\q}(\iwn) \\
 =  \int d\e_1 d\e_2\rho_2^{0}(\e_1,\e_2,\q)
F(\e_1,\iwn+\ivn)F(\e_2,\iwn)
\end{array}
\;\; ,
\end{equation}
and $F_{\k}(z) = F(\e(\k),z)$ is the lattice $f$-Green's function.
The matrices are indexed by Matsubara frequencies,
$\Pi_{ff}(\iwn,\iwm;\ivn) = \mat{\Pi}_{ff}|_{\iwn,\iwm}$, and
$\rho_2^{0}(\e_1,\e_2,\q)$ is the two particle density of 
states given by the underlying lattice\cite{MullerHartmann89}. Since
Eqn.~(\ref{equ-bethe-salp}) also holds  for the local two-particle
propagator $\mat{\Pi}_{ff}^{loc}$, we obtain the irreducible vertex
$\Gamma_{ff}$
and substitute it into (\ref{equ-bethe-salp}) to derive
\begin{equation}
\label{equ-ff-ph}
\mat{\Pi}_{ff}(\q) = 
\mat{\Pi}_{ff}^{loc}
%%\frac{1}{
\left[
\mat{1} - \frac{1}{\beta}\left[
[\mat{\chi}_{ff}^{0}]^{-1}_{loc}
-
[\mat{\chi}_{ff}^{0}(\q)]^{-1}
\right]
\mat{\Pi}_{ff}^{loc}
\right]^{-1}
%%}
\;\; ,
\end{equation}
which is valid for each Boson frequency $\ivn$
separately\cite{Jarrell95,Pruschke96}. In order to study possible phase
transitions, I restricted my calculations to the fully symmetric phase.
I used a Gaussian non-interacting
density of states $\rho_0(\e) = \frac{1}{t^*\sqrt{\pi}}
e^{-(\e/t^*)^2}$ for a simple cubic lattice in large
dimensions, and focused on the static $ff$-spin susceptibility
\begin{equation}
\label{equ-ff-sus}
%%\chi_{ff}(\ivn=0,\q) 
\chi_{ff}(\q) 
= \frac{1}{\beta^2}\sum_{\iwn,\iwm} \Pi_{ff}(\iwn,\iwm;0,\q)
e^{\delta(\iwn+\iwm)} \;\; .
\end{equation}
The diverges  of $\chi_{ff}(\q)$ originating from   zeros
of the denominator of (\ref{equ-ff-ph}) determine the phase
transitions, resembling a Stoner like expression for the lattice
susceptibility. I would like to mention that the  
convergence factors in Eqn.(\ref{equ-ff-sus}) are important for evaluating
the susceptibility analytically and {\em numerically}. 
The conduction electron susceptibility $\chi_{cc}$ and
the mixed terms $\chi_{cf}$ and $\chi_{fc}$ also contribute to the
total susceptibility. %%\cite{Grewe88}
However, since the unperturbed conduction electrons 
form a Fermi-gas, any irreducible interaction is generated by the
$f$-electrons\cite{Grewe87}.
Therefore, divergences 
occur collectively for all contributions when $\chi_{ff}(\q)$
diverges due to $c-f$ mixing. In large spatial dimensions,  
$\rho_2^{0}(\e_1,\e_2,\q)$ obtains its $q$  dependence
only through lattice dependent structure factor $\eta(\q)=\sum_i
\cos(q_i)/D$ in the case of  a hyper cubic lattice of
$D$-dimensions\cite{MullerHartmann89}.  

In context of Heavy-Fermion systems\cite{GreweSte91}, the terms
anti-ferromagnetism (AF) and spin density wave (SDW) are often used
interchangeably, since it is hard to distinguish experimentally  true
incommensurable band magnetism from complicated yet  commensurable
magnetic structures.
Here, I will call a transition
{\em anti-ferromagnetic} if the divergence occurs at $\eta(\q) = -1$,
{\em ferromagnetic}, if $\eta(\q) = 1$, while all
other cases will be labeled as {\em SDW}. 
To be specific, I choose the wave vector $\q =
q_0(1,1,\cdots)$ along the nesting direction instead of $\eta(\q)$
in all figures. 
I define the characteristic temperature of the lattice $T^*$ as the
temperature  at which the initial moment of the lower lying doublet
is screened down to 40\% \cite{SchillerAndCox98}. This energy scale is
the lattice analog to the single ion Kondo temperature. 
Taking into
account the  lack of universality due to the ongoing renormalization
of the conduction band for $T<T^*$, it is not surprising if $T^*$ does
not coincide with the energy scale of the electrical response $T_{FL}$, as
demonstrated in Ref.~\cite{Suellow99}. Hints are also seen
in calculations for low carrier concentrations\cite{JarrellNiki96}. 

%%%%%%%%%%%%%%%%%%%%%%%%%%%
%% results
%%%%%%%%%%%%%%%%%%%%%%%%%%%
\paragraph*{Results.}
The Anderson width $\Delta = V^2\pi \rho_0(0)$ is used throughout the
paper as the energy unit. All calculations have been performed for
a fixed band width $t^* = 10$  which yields  the hybridization
strength $V^2 = \Delta t^*/\sqrt{\pi}$.
If one defines the
effective dimensionless coupling constant $g=\rho_0(0) |J| = \rho_0(0)
V^2/|\e_f| = \Delta/(\pi |\e_f|)$, and requires, that $\Delta/\e_f| < 1$
as a criterion for excluding the 
intermediate valence regime, the TCPAM can only describe the weak
coupling regime ($g < 1/\pi$) of the two-channel Kondo lattice model.
The chemical 
potential $\mu$ has been varied from $-5$ to $5$ to obtain different
conduction electron occupancies $n_c$: at $n_c=4$ the c-bands are
completely filled.

The  occupancy of the magnetic doublet decreases monotonically with 
increasing $c$-band filling. Even though the effective DOS is altered
symmetrically around half filling, the different $f$-valent states
$|\sigma\!\!>$ ($|\alpha\!\!>$) involve admixture of virtual hole
(particle)  band states which are increased (decreased) with
decreasing chemical potential. Therefore, $T^*$ decreases more rapidly
with conduction band depletion than with conduction band
filling. Moreover, assuming $T^*\propto \exp{-1/(J\rho)}$, the change
of $T^*$ becomes more significant for decreasing $J$. This is exactly
what is found numerically of the occupancy dependence of $T^*(n_c)$
for different $\e_f$.  

\begin{figure}[htb]
\epsfxsize 80mm
%%\epsfbox{tc-tk-q-ef1.eps}
\epsfbox{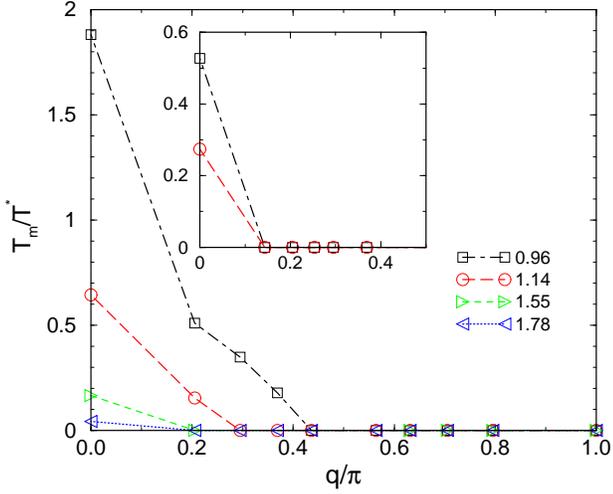}
\caption{Mixed-valent regime:
$T_m/T^*$ versus $q_0$ for different conduction electron
fillings. A ferromagnetic transition is found from $1/4$ to $1/2$
conduction band filling for $\e_f = -1\Delta$. The inset shows
$T_m/T^*$ for $\e_f =0$.}
\label{fig-1}
\end{figure}

Generally, magnetic phase transitions in the periodic Anderson model
are expected to be AF for the stable moment  or  SDW 
for the Kondo regime\cite{GreweSte91} 
stemming from inter and intra band excitations in a hybridized band
picture. This picture is not truly  applicable in the two channel model due
to lack of quasi-particles at the chemical potential. 

In the mixed valent (IV) regime ( $\e_f/\Delta = -1$) the largest
$T_m$ is found for a ferromagnetic  transition at less than  half band
filling (Fig.~\ref{fig-1}). With the exception of $n_c =1 $,
$T_m/T^*<1$. 
Independently, I also verified the existence of a ferromagnetic phase by
calculating the spin polarization in an
external magnetic field: its magnitude was decreased from a finite value
of the order of $10^{-4}T^*$ to zero field. 
A zero field magnetization was found only below $T_m$. Since the used
XNCA method works  reliably only in the multi-channel case
\cite{AndersJarCox97}, accurate determination of a magnetization curve
below $T_m$ was not possible. 

In the single channel PAM, no magnetism is expected in the IV regime due
weakness of the effective exchange couplings. In the two channel PAM,
however, incoherent scattering of the conduction electrons remains
present in the 
paramagnetic phase down to low temperatures. Energy can be gained by
reducing the scattering rate in  a spin-polarized state. This gain grows
 with increasing occupancy of the magnetic doublet. Therefore, the
transition is found at low $c$-carrier concentration, where charge
fluctuations mixes increasing number of $c$-hole states to local
magnetic ground state, stabilizing the local moment.
Because charge fluctuations are ignored  in the Kondo lattice model,
it has not been observed before \cite{JarrellPangCox97}. 
The FM is clearly a strong coupling effect,
since it is caused by high energy charge fluctuations \cite{CeSb}. It leads
to a higher mobility of the $c$-electrons measurable through an
enhancement of DOS in the majority spin above the chemical potential.
In the FM phase, the
increasing magnetization quenches the spin channel and the model
crosses over to a single channel PAM\cite{AndersJarCox97}. Below
$T_m$, Fermi-liquid properties are regained.

\begin{figure}[htb]
\epsfxsize 80mm
%%\epsfbox{tc-tk-two-fig.eps}
\epsfbox{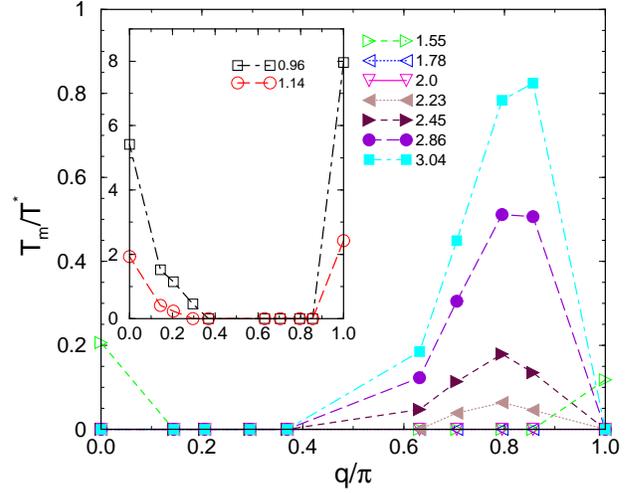}
\caption{
Kondo regime: $T_m/T^*$ versus $q_0$ for different conduction electron
fillings. $\e_f = -2\Delta$. The inset shows $T_m/T^*$  near quarter filling.
}
\label{fig-2}
\end{figure}

The picture changes significantly in the crossover regime,
$\e_f/\Delta = -2$ (Fig.~\ref{fig-2}): although the
occupancy changes moderately from $0.82$ to $0.74$
 with increasing band filling, the maximum of $T_m$ is now found  at a
finite $q$ vector 
$q_0 \approx  .85\pi$ indicating a SDW transitions for band fillings larger 
than 0.5. Around half filling, the Kondo effect  dominates. The
effective local moment is already quenched to 7\% 
of its high temperature value at  the SDW transition temperature
$T_m$. The transition occurs well below $T^*$ while with increasing
effective moment, the transition is shifted towards higher
temperatures. 
Near quarter filling, the local moments are only weakly screened: 
the effective moment $\mu_{eff}^2 = 0.6\mu_{free}^{2}$, the
ratio  $T_m/T^*$ is very large (inset of Fig.~\ref{fig-2}) and an AF
transition is observed similar to the stable moment situation discussed next.

\begin{figure}[htb]
\epsfxsize 80mm
\epsfbox{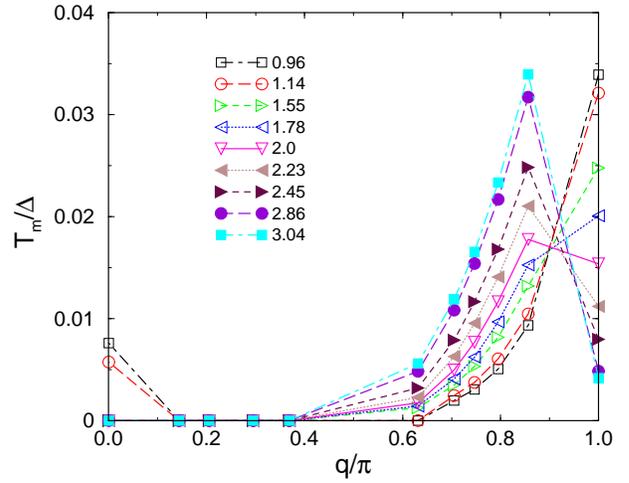}
\caption{Stable moment regime: 
$T_m/\Delta$ versus $q_0$ for different conduction electron
fillings. $\e_f = -3.5\Delta$.
}
\label{fig-3}
\end{figure}

Lowering $\e_f$ further, we arrive at the stable local moment regime
(Fig.~\ref{fig-3}). For $n_c < 2$, an AF transition is found. 
Note that $T_m$ is not rescaled with $T^*$ 
in this case, indicating that the screening of the moments is very low
(about 25\%).
Increasing the band filling,  the admixture of $|\alpha\!\!>$
states into the ground state increases, which decreases the local moments: for
$n_c>2$ SDW transitions at large $q$ vectors are found.
For positive $\e_f$, no magnetic phase transition is found. 

%%%%%%%%%%%%%%%%%%%%%%%%%%%
%% Discussion
%%%%%%%%%%%%%%%%%%%%%%%%%%%

\paragraph*{Discussion.}
I have established, for the first time, a ferromagnetic solution of
the two-channel periodic Anderson model in the intermediate valence
regime for low carrier concentration.
Energy and mobility  is gained by reducing the incoherent scattering of the
conduction electrons in a spin-polarized state. The loss of energy due 
to spin polarization of the conduction band is low for small  band
fillings. The transition 
is driven by high energy charge fluctuations indicating the strong
coupling nature of the FM transition. 
The presences of FM transitions and strong ferromagnetic fluctuations
suggest the possibility of triplet superconductivity. Preliminary
studies indicate a divergence of the pair susceptibility in the
spin/channel triplet sector, implying an odd-frequency gap function.

In the Kondo regime, I find SDW transitions, with a large wave vector
$q$, for band fillings larger than 1/2. Additionally, nearest
neighbour AF and tendency towards ferromagnetism is observed for low
carrier concentrations. In the stable moment regime AF transitions are
found at less than half filling  and SDW transition above half filling.
However, I have excluded a detailed treatment of
band-structure effects in finite dimensions within DMFT, which can be
included through Equ.~(\ref{equ-3}). The bare two-particle density of
states then becomes truly $\q$-dependent. I hope to clarify its
influence on the spin susceptibility in three dimensions in the near
future.%% \cite{steffen-prelim}.

In all the cases I investigated, I found $T_m 
\propto E_{RKKY} = \mu_{eff}^2< \!\! X_\sigma\!\!>^2J^2\rho_0(\mu)$. 
where $<\!\!X_\sigma\!\!>$ is the occupation number of the magnetic
ground state doublet of the each ion. The ratio $E_{RKKY}/T_m$
decreases monotonically from around 3. to 1.5 with decreasing
effective local moment for $\e_f=-2$. The magnetic transition is
{\em always} governed by an effective RKKY energy $E_{RKKY}$, which takes
into account the screening of the local moments. $E_{RKKY}$ can be  
one order of magnitude smaller than the bare RKKY energy.

Experimentally, the change of parameters in real materials is much more
difficult: usually either the size of the unit cell is changed through
isostructural substitution or through external pressure. 
Essentially, the overlap between nearest neighbours is varied which
yields a change  in the hybridization $V$ and the bandwidth $t^*$. 
It is believed, that the core energy difference $\e_f$ stays unaltered
\cite{JRoehler87}. Under the assumption, that only hybridization and
bandwidth are altered by pressure, 
application of positive pressure will be equivalent to a reduction of
$|\e_f|$. The results presented show clearly the same tendency as observed in
experiments \cite{Suellow99}:  the application of pressure leads to a suppression of
magnetic phase transitions. Local moments are destroyed through
hybridization induced  de-localization of the f-electrons.

The author is very thankful for fruitful discussions with
Dr.~Th.~Pruschke and Prof.~N.~Grewe, who also carefully read the
manuscript. The work was in part  funded by the 
Sonderforschungsbereich 
252, {\em Elektronisch hochkorrelierte metallische Materialien}. All
calculation have been performed on a 8 CPU LINUX cluster.

%%%%%%%%%%%%%%%%%%%%%%%%%%%
%% References
%%%%%%%%%%%%%%%%%%%%%%%%%%%


\begin{thebibliography}{99}
\bibitem{CeSb}
Since the mechnism appears to
be generic, it would be worthwhile to investigate, whether the FM
transitions in CeSb can be understand along the same lines.
\end{thebibliography}
\end{document}